\documentclass{aa}
\usepackage{txfonts}
\usepackage{graphicx, graphics}
\usepackage{CJK}
\usepackage{bbm}
\usepackage{amsmath, amssymb, bm}
\usepackage[nodayofweek]{datetime}
\input{prepcitationformat.txt}
\usepackage{txfonts}
\usepackage{algpseudocode}
\usepackage{algorithm}

\usepackage{mathtools}

\DeclarePairedDelimiter\ket{\lvert}{\rangle}
\DeclarePairedDelimiterX\braket[2]{\langle}{\rangle}{#1 \delimsize\vert #2}

\usepackage[switch]{lineno} 

\usepackage{tikz}

\definecolor{lime}{HTML}{A6CE39}
\DeclareRobustCommand{\orcidicon}{%
    \raisebox{-3pt}{\begin{tikzpicture}
    \filldraw [lime, yshift=-2pt] (0, 0) circle [radius=0.16]
    node[white] {\raisebox{1pt}{\hspace{0.5pt}\fontfamily{qag}\selectfont\tiny i\scalebox{0.8}{D}}};
    \end{tikzpicture}}
    \hspace{-2.5mm}
    \vspace{-0.25pt}
}

\newdateformat{monthyearday}{
  \THEYEAR\ \monthname[\THEMONTH] \twodigit{\THEDAY}}
\newdateformat{daymonthyear}{%
  \twodigit{\THEDAY}\ \monthname[\THEMONTH] \THEYEAR}
\newcommand\numberthis{\addtocounter{equation}{1}\tag{\theequation}}

\newcommand{\orcidauthor}[2]{#2\href{http://orcid.org/#1}{\orcidicon}}

\titlerunning{DIKL for High Contrast Imaging}
\authorrunning{Ren}

\begin{document}
\begin{CJK*}{UTF8}{gbsn}
\title{Karhunen--Lo\`eve Data Imputation in High Contrast Imaging\thanks{FITS images are only available at the CDS via anonymous ftp to \url{cdsarc.cds.unistra.fr} (\url{130.79.128.5}) or via \url{https://cdsarc.cds.unistra.fr/viz-bin/cat/J/A+A/}}}

\author{
\orcidauthor{0000-0003-1698-9696}{Bin B. Ren (任彬)}\thanks{Marie Sk\l odowska-Curie Fellow}
}

\institute{
Universit\'{e} C\^{o}te d'Azur, Observatoire de la C\^{o}te d'Azur, CNRS, Laboratoire Lagrange, Bd de l'Observatoire, CS 34229, 06304 Nice cedex 4, France; \email{\url{bin.ren@oca.eu}}
}

\date{Received 04 July 2023 / Revised 30 August 2023 / Accepted 31 August 2023} 

\abstract
{Detection and characterization of extended structures is a crucial goal in high contrast imaging. However, these structures face challenges in data reduction, leading to over-subtraction from speckles and self-subtraction with most existing methods. Iterative post-processing methods offer promising results, but their integration into existing pipelines is hindered by selective algorithms, high computational cost, and algorithmic regularization. To address this for reference differential imaging (RDI), here we propose the data imputation concept to Karhunen--Lo\`eve transform (DIKL) by modifying two steps in the standard Karhunen--Lo\`eve image projection (KLIP) method. Specifically, we partition an image to two matrices: an anchor matrix which focuses only on the speckles to obtain the DIKL coefficients, and a boat matrix which focuses on the regions of astrophysical interest for speckle removal using DIKL components. As an analytical approach, DIKL achieves high-quality results with significantly reduced computational cost (${\sim}3$ orders of magnitude less than iterative methods). Being a derivative method of KLIP, DIKL is seamlessly integrable into high contrast imaging pipelines for RDI observations.}


\maketitle


\section{Introduction}
High contrast imaging aims at detecting and characterizing faint signals surrounding bright central sources \citep[e.g.,][]{oppenheimer09, benisty23, currie23}. To reach this goal, coronagraphs are used to suppress central light, and post-processing -- combined with observational techniques -- is implemented to remove residual light and speckles \citep[e.g.,][]{pueyo18, follette23}. With these setups, existing observational techniques and methods are efficiently detecting and characterizing point sources such as planets and brown dwarfs \citep[e.g.,][]{nielsen19, vigan21}, as well as circumstellar disks and quasar host galaxies in polarized light \citep[e.g.,][]{benisty23, gratadour15}. Total intensity detection for extended structures, however, is still prone to data reduction artifacts with most existing methods for ground-based observations \citep[e.g.,][]{ruane19, xie22}. Their characterization is limited by compromised data quality \citep[e.g.,][]{milli12, mazoyer20, olofsson23, xie23}.

To detect and characterize extended structures in total intensity, several statistics-based methods and their derivative methods are being used  \citep[e.g.,][]{lafreniere07, soummer12, amara12, ren18, ren20, pairet18, pairet21, flasseur21, samland21, juillard22, berdeu22}. Nevertheless, these methods are either prone to severe overfitting that can alter the morphology and surface brightness of such structures \citep[e.g.,][]{lafreniere07, soummer12}, or being highly selective on reference regions and computationally intensive \citep[e.g.,][]{ren18, ren20}, or subject to regularization terms that may provide uncertain results on the morphology and surface brightness of extended structures \citep[e.g.,][]{flasseur21, pairet21, juillard22}. Meanwhile, a careful selection of data reduction regions \citep[e.g.,][]{galicher11, milli12, milli17, perrin15} could provide better results than non-statistical methods, yet these regions need to be adjusted for different systems, and such a selection is challenging when the disk morphology is face-on, complex, or even unknown. Using non-negative matrix factorization \citep[NMF:][]{lee01}, \citet{ren20} showed that data imputation with sequential NMF (DI-sNMF) could be a promising statistical method for simple disk morphology, and it can provide high-quality results with theoretically minimized signal alteration. However, the data imputation concept has only been applied in \citet{ren20} for NMF in high contrast imaging, and it can provide high-quality results only when some strict requirements are satisfied \citep[e.g.,][]{ren20, olofsson23, xie23}. What is more, DI-sNMF is computationally inefficient due to its iterative nature \citep[e.g.,][]{ren18}, calling for other analytical methods that can implement the data imputation concept.

In high contrast imaging, the principal-component-analysis-based Karhunen--Lo\`eve Image Projection \citep[KLIP:][]{soummer12, amara12} method is one of the analytical method standards for data reduction. KLIP decomposes reference images into an orthogonal basis through Karhunen--Lo\`eve (KL) transform, and it has been included in post-processing pipelines \citep[e.g.,][]{pyklip, vip, pynpoint, adijl}. It is widely used by surveys on both exoplanets \citep[e.g.,][]{nielsen19, vigan21} and circumstellar disks \citep[e.g.,][]{esposito20, xie22, cugno23, wallack23}. Due to its over-fitting nature \citep[e.g.,][]{soummer12, pueyo16}, KLIP encounters difficulty in detecting and characterizing complex and extended structures. While iterative KLIP derivatives \citep[e.g.][]{pairet18, ginski21, stapper22} might help alleviate the over-subtraction and self-subtraction in certain observational setups, there are still persistent residual signals and self-subtraction artefacts that pose challenges to data interpretation. As a data replacement (i.e., imputation) attempt with KLIP, \citet{hunziker18} and \citet{xuan18} used KLIP-based background subtraction in an interpolation approach: first perform KL decomposition, then conduct KLIP while masking out the central regions. However, due to the masking of central regions, the KLIP procedure within were performed on a non-orthogonal basis. Such a KLIP treatment can introduce over-fitting, which is manifested as a higher empirical model for the background, and yields negative surrounding regions even before post-processing. Noticing this over-fitting, \citet{xie23} thus used DI-sNMF for background removal and post-processing to extract a double-spiraled system to enable precise spiral motion measurement. After all, a proper data imputation with KLIP would offer a promising way to produce high-quality results with a high computational efficiency.

In the upcoming era of Extremely Large Telescopes (ELTs), the reference differential imaging (RDI) observational setup, including star-hopping \citep{wahhaj21}, could likely dominate the observation modes to produce high-quality images of exoplanets, circumstellar disks, and quasar host galaxies  (hereafter ``astrophysical signals'') in total intensity. In an RDI observation, a reference star (or multiple reference stars), which does not host circumstellar signals, is used to empirically capture the point spread function and speckles of the target star or quasar to reveal the target's astrophysical surroundings. With large collecting areas \citep[e.g.,][]{elt}, ELTs can reach high signal-to-noise ratios in a significantly smaller amount of time \citep[e.g.,][]{bowens21} than existing telescopes. In comparison, it might be suboptimal for the classical angular differential imaging \citep[ADI:][]{marois06} observation setup to be deployed on ELTs, since ADI requires a sufficient amount of field rotation of the sky for data processing \citep[e.g.,][]{milli12, xuan18}, yet sky rotation rate is a natural phenomenon that cannot be changed. To reduce and process the ELT data with high fidelity in RDI observations, iteratively methods could be potentially offer the best results. However, these iterative methods might be limited due to their significantly high computational cost than analytical methods, especially when there are high-resolution observations with a large number of datasets. Therefore, an analytical RDI data reduction method is needed to provide high-quality results in an efficient way.

In this study, by engineering the mathematical basics for KLIP, we enable it with the data imputation concept (DIKL). By modifying two steps in the standard KLIP procedure, and illustrating the DIKL results, we expect DIKL to be a promising method in RDI data reduction. In comparison with the established iterative DI-sNMF method which offers high-quality RDI results \citep[e.g.,][]{olofsson23, xie23}, such modifications allow an analytical data imputation by DIKL. It is computationally more efficient than iterative methods such as DI-sNMF by ${\sim}3$ orders of magnitude, and it can provide results with comparable quality as DI-sNMF in RDI datasets.

\section{Method}\label{sec-obs}
In high contrast imaging observations which require the usage of coronagraphs and adaptive optics systems, light signals from a central object (e.g., star, quasar) received on the detectors are superimposed on the astrophysical signals \citep[e.g.,][]{pueyo18, follette23}. The goal of post-processing is to extract the faint astrophysical signals (e.g., exoplanets, circumstellar disk, host galaxy) that are hidden behind these bright signals of non-astrophysical origin (e.g., adaptive optics halo, speckles, thermal background). For simplicity, we refer all these non-astrophysical signals as ``speckles'' hereafter. For an image which contains astrophysical signals, we refer it as target $t$. For a set of images that only contain speckle signals, we refer them as reference images $R$.\footnote{A target can serve as a reference in scenarios such as ADI, in which astrophysical signals move with respect to speckles in different images.} In post-processing, we use the reference images to obtain the representative features of the speckles, then remove these features from a target image to reveal astrophysical signals in the residuals.

\subsection{Overview of KLIP}
For a reference image array containing $n_{\rm ref}\in\mathbb{N}$ reference images each with $n_{\rm pix}\in\mathbb{N}$ pixels, we denote the array as $R\in\mathbb{R}^{n_{\rm pix}\times n_{\rm ref}}$. A column in $R$ contains a reference image that is converted to a one-dimensional column vector (e.g., concatenating all columns of one image). We note that the notations here follow that of statisticians and computer scientists, instead of existing publications by astronomers \citep[e.g.,][]{soummer12, pueyo16}, to enable an efficient delivery of messages to non-astronomy readers in coding. Specifically, we have the reference array $R = \left[R_1, \cdots, R_{n_{\rm ref}}\right]$, with $R_i = (R_{i1}, \cdots, R_{in_{\rm pix}})^\top$ for $i\in\left\{1, \cdots, n_{\rm ref}\right\}$ and $\top$ for matrix transpose, where $R_{ij}\in\mathbb{R}$ for $j\in\{1, \cdots, n_{\rm pix}\}$. 

For a target image, we can convert it to a column vector $t\in\mathbb{R}^{n_{\rm pix}\times1}$. To remove the speckles from the target image, a statistics-based post-processing method first transforms the reference array $R$ to obtain a basis. This basis contains the representative features of the reference array (i.e., components). We then remove the representative features from the target vector. After post-processing, astrophysical signals are expected to reside in the residual image from feature removal.

The KLIP algorithm in \citet{soummer12} is consisted of two steps: the Karhunen--Lo\`eve (KL) transform of the reference matrix $R$ to obtain the KL component basis, and the projection of the target vector $t$ on the KL basis. For KLIP, the columns in both $R$ and $t$ are zero-spatial-mean. To reveal the astrophysical signals, the KLIP projection is then removed from the target image. To distinguish the difference between KLIP and DIKL later in this work, here we denote the reference array for KLIP as matrix $R^{(a)}\in\mathbb{R}^{n^{(a)}_{\rm pix}\times n_{\rm ref}}$, with $n^{(a)}_{\rm pix} \leq n_{\rm pix}$ allowing for the selection of smaller regions in reference images. Similarly, we denote the target vector as $t^{(a)}\in\mathbb{R}^{n^{(a)}_{\rm pix}\times 1}$ in KLIP processing. With these notations, we review the KLIP procedure as follows.

 \begin{figure*}[htb!]
\centering
 	\includegraphics[width=\textwidth]{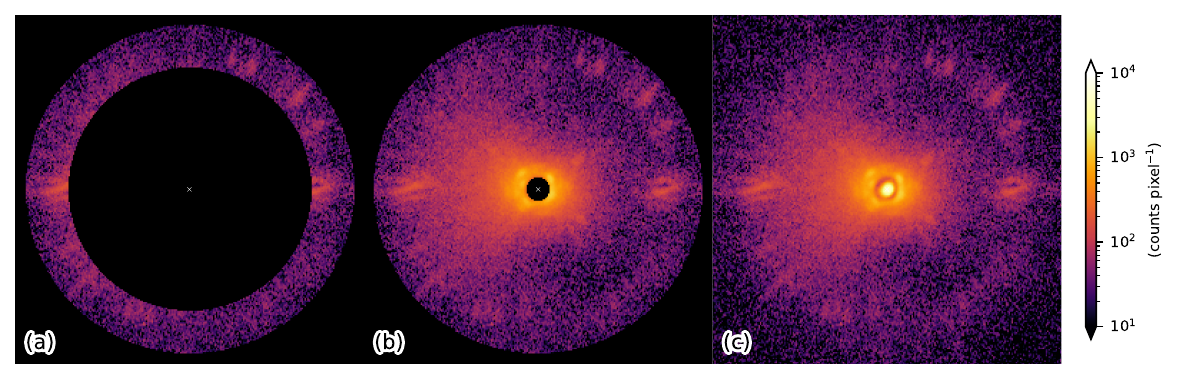}
     \caption{Partitioning of an image for DIKL. (a) An anchor image used to construct KL basis. (b) A boat image used to construct DIKL basis. (c) A complete image without partitioning. Note: the images here are vectorized to constitute the columns of the $R^{(a)}$, $R^{(b)}$, and $R$ matrices, respectively.}
     \label{fig-abc}    
 \end{figure*} 
 
In the KL transform step, for a reference array $R^{(a)}$, its covariance matrix $\Sigma^{(a)} = R^{(a)\top}R^{(a)} \in\mathbb{R}^{n^{(a)}_{\rm pix}\times n^{(a)}_{\rm pix}}$ is real symmetric. The spectral decomposition of $\Sigma^{(a)}$ is
\begin{equation}\label{eq-decomp}
\Sigma^{(a)} = R^{(a)\top}R^{(a)} = Q^{(a)}D^{(a)}Q^{(a)\top},
\end{equation}
where $D^{(a)} = {\rm diag}\{\lambda_1^{(a)}, \cdots, \lambda_{n_{\rm ref}}^{(a)}\}$ is a diagonal matrix whose diagonal entries are the eigenvalues of $\Sigma^{(a)}$ with $\lambda_k^{(a)}\geq\lambda_{k+1}^{(a)}>0$, and $Q^{(a)}\in\mathbb{R}^{n_{\rm ref}\times n_{\rm ref}}$ is an orthonormal matrix with its columns being the corresponding eigenvectors with $Q^{(a)\top}Q^{(a)}=Q^{(a)}Q^{(a)\top}=I$. Left multiply Eq.~\eqref{eq-decomp} by $Q^{(a)\top}$ and right multiply by $Q^{(a)}$, we have 
\begin{equation*}
[R^{(a)}Q^{(a)}]^\top R^{(a)}Q^{(a)} = Q^{(a)\top} R^{(a)\top}R^{(a)}Q^{(a)} = D^{(a)},
\end{equation*}
where we now have diagonalized the covariance matrix. By defining
\begin{equation}\label{eq-kl}
Z^{(a)}= R^{(a)}Q^{(a)},
\end{equation}
we have an orthogonal KL component basis $Z^{(a)}\in\mathbb{R}^{n^{(a)}_{\rm pix}\times n_{\rm ref}}$. The columns of the KL basis can be divided by the corresponding square root of the eigenvalues \citep[e.g.,][]{soummer12}, and thus creating an orthonormal basis $Z^{(a)}$. While we do not adopt this normalization since it does not impact the following target modeling procedure in the datasets used later in this study, such a normalization \citep[e.g., \texttt{VIP}:][]{vip} might be necessary for other datasets.

In the KL projection step, for a given target image vector with zero-spatial-mean $t^{(a)}\in\mathbb{R}^{n^{(a)}_{\rm pix} \times 1}$, we can project and remove the KL component to obtain the residual $r^{(a)}\in\mathbb{R}^{n^{(a)}_{\rm pix} \times 1}$,
\begin{equation}\label{eq-klip}
r^{(a)} = t^{(a)} - \sum_{k=1}^{K_{\rm klip}} c^{(a)}_k Z^{(a)}_k,
\end{equation}
where $K_{\rm klip}\in\{1, \cdots, n_{\rm ref}\}$ is the cutoff count of the KL components \citep{soummer12}, and 
\begin{equation}\label{eq-kl-coef}
c^{(a)}_k = t^{(a)\top} Z^{(a)}_k
\end{equation} is a scalar with $c_k^{(a)}\in\mathbb{R}$, which denotes the projection from the target onto the $k$-th KL component of $R^{(a)}$.\\

 In the next step of this study, we will modify the KLIP procedure to introduce DIKL post-processing. To perform DIKL data reduction, we need to focus on different regions of the reference array $R$. For the full reference array $R$, we can reorder or duplicate certain of its rows to obtain two submatrices,
\begin{equation}
\left\{\begin{matrix}
R^{(a)} = S^{(a)}R\\ 
R^{(b)} = S^{(b)}R
\end{matrix}\right.,\label{eq-rab}
\end{equation}
where $R^{(a)}\in\mathbb{R}^{n^{(a)}_{\rm pix}\times n_{\rm ref}}$, $R^{(b)}\in\mathbb{R}^{n^{(b)}_{\rm pix}\times n_{\rm ref}}$, with $n^{(a)}_{\rm pix} \leq n_{\rm pix}$ and $n^{(b)}_{\rm pix} \leq n_{\rm pix}$. Selection matrices $S^{(a)}\in\mathbb{B}^{n^{(a)}_{\rm pix}\times n_{\rm pix}}$ and $S^{(b)}\in\mathbb{B}^{n^{(b)}_{\rm pix}\times n_{\rm pix}}$ are boolean, and they are used to select specific rows in $R$ to form $R^{(a)}$ and $R^{(b)}$, respectively. For example, to select the $j$-th row from $R$ and store it in the $i$-th row of $R^{(a)}$, we have $S^{(a)}_{ij}=1$ (and $S^{(a)}_{ij}=0$ otherwise).  To be more informative on the naming conventions of the superscripts in Eq.~\eqref{eq-rab}, we refer the two selected matrices as an anchor matrix $R^{(a)}$ and a boat matrix $R^{(b)}$. The anchor matrix $R^{(a)}$ covers the regions which only host speckle signals, and the boat matrix $R^{(b)}$ can host astrophysical signals (and it can also contain the anchor matrix, which is adopted in this study), see Fig.~\ref{fig-abc} for an illustration of the selected regions. 

\subsection{DIKL for Reference Differential Imaging}\label{sec-dikl}

The DIKL method modifies KLIP in two aspects. Specifically, the DIKL components are constructed by applying the eigenvectors from the anchor matrix $R^{(a)}$ to the boat matrix $R^{(b)}$, and the DIKL projection adopts the projection coefficient between anchor target $t^{(a)}$ and anchor KL basis $Z^{(a)}$.

On the one hand, DIKL modifies the KL transform in Eq.~\eqref{eq-kl}. The DIKL basis follows
\begin{equation}\label{eq-kl-transfer}
{Z'}^{(b)}= R^{(b)}Q^{(a)},
\end{equation}
where we now instead use the eigenvectors from $R^{(a)}$ to construct the DIKL basis for $R^{(b)}$ regions. Here we use the prime symbol for the DIKL basis to distinguish it from a normal KL basis for the reference matrix $R^{(b)}$. In comparison, the KL basis for $R^{(b)}$ is ${Z}^{(b)}$, and it is not calculated nor used in this study.

On the other hand, we modify the KL projection in Eq.~\eqref{eq-klip}. For a target vector $t^{(b)}\in\mathbb{R}^{n^{(b)}_{\rm pix} \times 1}$ with zero-spatial-mean, DIKL uses the coefficients from Eq.~\eqref{eq-kl-coef} between $t^{(a)}$ and $Z^{(a)}$, then applies the coefficients to the DIKL basis in Eq.~\eqref{eq-kl-transfer}. Specifically, the residual image after DIKL projection is
\begin{align*}
r^{(b)} &= t^{(b)} - \sum_{k=1}^{K_{\rm klip}} c^{(a)}_k {Z'}^{(b)}_k\\
	&= t^{(b)} - \sum_{k=1}^{K_{\rm klip}}  t^{(a)\top} Z^{(a)}_k {Z'}^{(b)}_k  \\
	&= \ket{S^{(b)}t} - \sum_{k=1}^{K_{\rm klip}} \braket{S^{(a)}t}{Z^{(a)}_k} \ket{{Z'}^{(b)}_k}, \numberthis \label{eq-DIKL}
\end{align*}
where the last equation adopts the bra--ket convention for the purpose of demonstration and connection with existing studies \citep[e.g.,][]{soummer12}.

With the above procedure, Eqs.~\eqref{eq-kl-transfer} and \eqref{eq-DIKL} establish the DIKL post-processing procedure, see Appendix~\ref{sec-app-pseudo} for a corresponding pseudocode and implementation instructions.  On the one hand, there is no direct projection of $t^{(b)}$ onto the KL basis of $R^{(b)}$, which ensures the non-overfitting of non-speckle signals for DIKL. On the other hand, the projection coefficient is obtained from where speckle-only signals are expected (in both the reference images and the target image). Both ensure that astrophysical signals in the boat regions of $t^{(b)}$ are not captured in the coefficients in Eq.~\eqref{eq-DIKL}, and thus it serves as the data imputation step for DIKL. In fact, DIKL only performs KL transform to the speckles in the anchor matrix in this work, which avoids the high variation of signals in the boat matrix (e.g., regions that are close to the coronagraph) to dominate the classical KL transform (which is performed on the entire image), see Fig.~\ref{fig-abc}.

We note that, however, Eq.~\eqref{eq-kl-transfer} is limited in the reduction of non-RDI data. For example, in ADI datasets, we need to mask out different pixels in an observation sequence \citep[e.g.,][]{milli17, ren20} to focus on speckle-only signals, since angular diversity is used to capture the speckles in an ADI observation sequence. We can indeed calculate an element in the covariance matrix by focusing only on the pixels that are not masked out in a pair of reference images, and perform spectral decomposition in Eq.~\eqref{eq-decomp} to obtain the corresponding eigenvectors. However, we cannot yet construct a basis following Eq.~\eqref{eq-kl} for KLIP, nor Eq.~\eqref{eq-kl-transfer} for DIKL, since weights in the eigenvector matrix $Q$ are assigned to data that are masked out (i.e., artificially ``missing'') in the reference array $R$. To apply DIKL to ADI datasets, we can interpolate the speckle-only signals for the masked out data \citep[e.g.,][]{perrin15} or iteratively fill the missing data \citep[e.g.,][]{bailey12} in $R$ for Eq.~\eqref{eq-kl-transfer}, yet such treatments beyond the current scope of this analytical study.

\section{Application}\label{sec-ana}
To demonstrate the DIKL method using on-sky RDI data, we retrieved datasets from the SPHERE instrument at the Very Large Telescope (VLT). We adopted data from IRDIS \citep{dohlen08} of SPHERE in the star-hopping mode \citep{wahhaj21}. The star-hopping mode hops between a target star and its reference star during an observation sequence, thus reaching a quasi-simultaneous capture of speckle change. We retrieved the UT 2021-09-06, UT 2022-02-07, and UT 2022-04-01 observations of circumstellar disk hosts -- HD~169142, PDS~201, and HD~129590 -- as well as their reference stars from programs 105.209E and 105.20GP. We preprocess the data using the \texttt{IRDAP} pipeline \citep{irdap1, irdap2}, and the stars are then located at the center of the images. In the IRDIS data, with each IRDIS pixel having a spatial scale of $12.25$~mas \citep{maire16}, the control ring -- interior to which the adaptive optics system can perform optimal dark hole correction to reveal faint objects -- spans between $85$~pixel and $115$~pixel from the matrix centers. For further processing, we masked out the matrix elements that are interior to a radius of $8$~pixel from the center. We display the partitioning of an image in Fig.~\ref{fig-abc} for and illustration of the anchor region and the boat region in this study. 

For the three groups of IRDIS datasets, the images for post-processing are the central $350\times350$ pixel from the \texttt{IRDAP} preprocessed data. For one exposure, there are two channels from the preprocessed data, and we added them to produce one image. There are $128$, $32$, and $32$ target images, with $36$, $12$, and $32$ corresponding reference images, for HD~169142, PDS~201, and HD~129590, respectively. To convert a 2-dimensional image to a column in a matrix for this study, in practice, we created a 2-dimensional binary mask whose entries are 1 when the corresponding pixels are selected in Fig.~\ref{fig-abc}(a)(b). For one image, we used the \texttt{where} function in \texttt{numpy} \citep{numpy} to select the pixels in the images, and created a column for the selected matrix. By performing this for all reference images, we created the selected anchor matrix $R^{(a)}$ or the boat matrix $R^{(b)}$ for further post-processing. By performing this once for a target image, we created the selected anchor target vector $t^{(a)}$ or the boat target vector $t^{(b)}$.

In post-processing, for the three IRDIS datasets that are in matrix form, we first followed Eq.~\eqref{eq-decomp} to perform spectral decomposition of the matrix elements in the anchor matrix (i.e., the IRDIS control ring). We then followed Eq.~\eqref{eq-kl-transfer} to conduct DIKL transform on the boat matrix -- the entire image (or the matrix elements interior to the outer edge of the control ring for the HD~129590 data) -- using the eigenvectors from the covariance matrix of the anchor references.  At last, to remove the speckles for a target image, we followed Eq.~\eqref{eq-DIKL} to perform DIKL reduction, where we adopted the KLIP projection coefficients from Eq.~\eqref{eq-kl-coef} between the anchor target and anchor references. We reshaped the 1-dimensional vectors to 2-dimensional image, then derotated the reduction results for all target images to north-up and east-left according to their corresponding parallactic angles calculated using \texttt{Pynpoint} \citep{pynpoint}. We calculated the element-wise median of the derotated reduction results as the combined image. We adopt the median-subtracted combined image as the final result.

\subsection{Residual Variance}
\begin{figure}[htb!]
\centering
 	\includegraphics[width=0.45\textwidth]{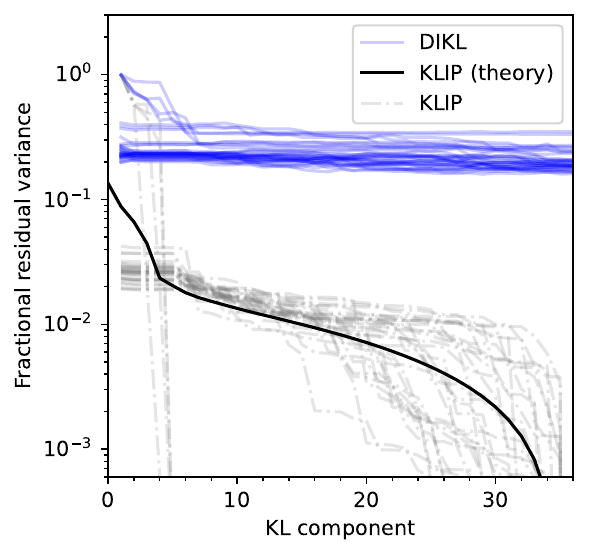}
     \caption{Fractional residual variance as a function of KL components for RDI. DIKL reaches higher plateaus than KLIP, illustrating the ability of information retention using DIKL.\\ (The data used to create this figure are available in the ancillary folder on arXiv.)}     \label{fig-frv}    
 \end{figure}

Variance of residual images can be informative of the existence of astrophysical signals. By performing DIKL on the reference images of the HD~169142 dataset, we generated the corresponding fractional residual variance (FRV) curves \citep[e.g.,][]{soummer12, ren18}. Specifically, for a reduced reference image, we divide its variance by that of the corresponding original image to generate the FRV. The FRV curves are then the FRV dependence as a function of KL components. Similarly, we generated the FRV curves for KLIP for comparison. For KLIP, FRV curves are expected to follow the fractional residual eigenvalues \citep{soummer12}, as can be seen in Fig.~\ref{fig-frv}.

 \begin{figure}[htb!]
\centering
 	\includegraphics[width=0.5\textwidth]{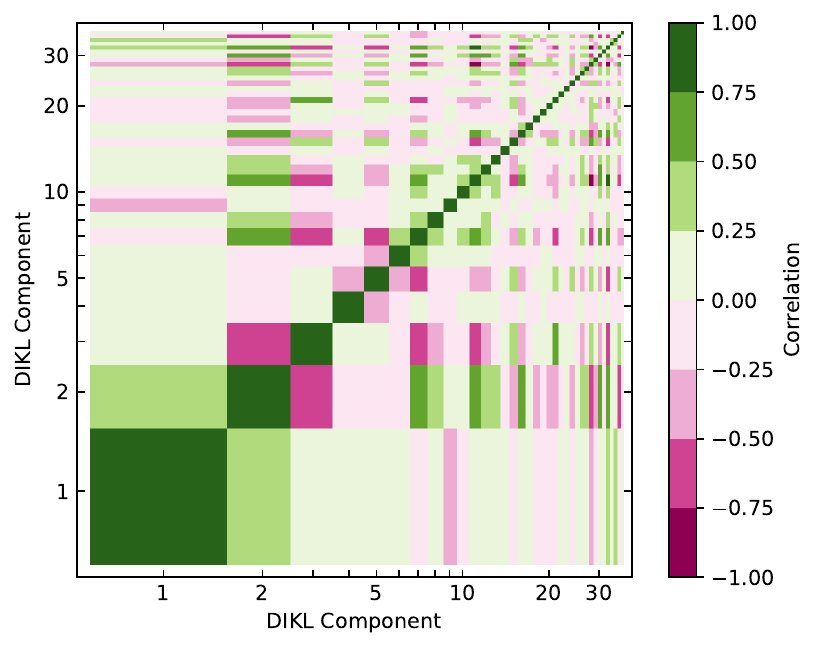}
     \caption{Correlation matrix of DIKL components. While KL components are mutually orthogonal, most DIKL components are not. However, with the FRVs in Fig.~\ref{fig-frv} suggesting that only the first ${\sim}5$ DIKL components contributes to the removal of speckles, the crosstalk from higher order components therefore does not impact DIKL data reduction. \\ (The data used to create this figure are available in the ancillary folder on arXiv.)}     \label{fig-corr-dikl}    
 \end{figure}

The FRV curves of DIKL converge to higher plateaus than those from KLIP. This convergence illustrates the less aggressive overfitting (or potentially non-overfitting of data; e.g., NMF: \citealp{ren18}) of DIKL. Similar as NMF, the FRV plateaus with DIKL is the indirect evidence that DIKL can retain more information of non-speckle signals than KLIP. In addition, the FRVs for DIKL reaches to stable values with ${\approx}5$ DIKL components, suggesting that the rest of the components have negligible contribution in the DIKL process.

Due to the non-orthogonality of a DIKL basis, we witnessed a minor level of overfitting of the data, and thus we subtracted the median of the DIKL residuals to manually offset this effect. We present in Fig.~\ref{fig-corr-dikl} the correlation matrix for DIKL components: there exist crosstalks in the form of non-zero off-diagonal elements. However, given that the contribution of higher order DIKL components (${\gtrsim}5$) are negligible in the FRVs in Fig.~\ref{fig-frv}, the DIKL crosstalk only impacts the first few components, and the crosstalk does not contribute to data reduction beyond the first ${\approx}5$ DIKL components due to the reaching of FRV plateaus in Fig.~\ref{fig-frv}.

\subsection{DIKL Imaging}

We used DIKL to reduce the star-hopping RDI observations of HD~169142, PDS~201, and HD~129590, which are known circumstellar disk hosts. The three disks have inclinations from nearly face-on to roughly edge-on \citep[e.g.,][]{pohl17, wagner20, matthews17, olofsson23}. We also reduced the datasets with KLIP and DI-sNMF for comparison.

\begin{figure}[htb!]
\centering
 	\includegraphics[width=0.45\textwidth]{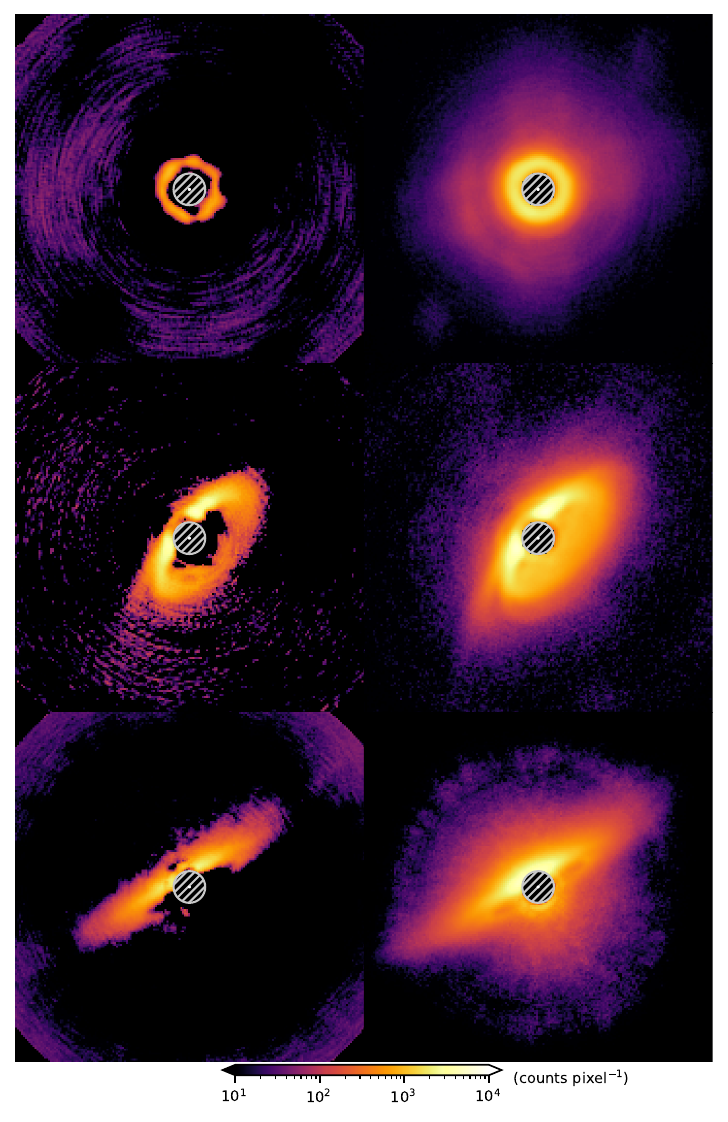}
     \caption{Comparison of reduction results using KLIP (left) and DIKL (right) for nearly face-on to roughly edge-on systems. For HD~169142 (left), PDS~201 (middle), and HD~129590 (right), DIKL reaches data quality that might be comparable with polarized light observations.\\ (The data used to create this figure are available in the ancillary folder on arXiv.)}
     \label{fig-compare}    
 \end{figure}

Between the DIKL and KLIP results  in Fig.~\ref{fig-compare}, for the nearly face-on HD~169142, DIKL retrieves a two-ringed system, KLIP can nevertheless only recover the inner ring that is close to the coronagraph with compromised data quality. The DIKL result resembles the disk image obtained in polarized light, which thus demonstrates its superiority over KLIP in conserving face-on structures. Similarly, the PDS~201 and HD~129590 results have fine extended structures only seen in DIKL when compared with KLIP. In comparison, KLIP removes signals from the disks, altering disk morphology that poses challenges in data interpretation \citep[e.g.,][]{wagner20, olofsson23}.

\begin{figure}[htb!]
\centering
 	\includegraphics[width=0.5\textwidth]{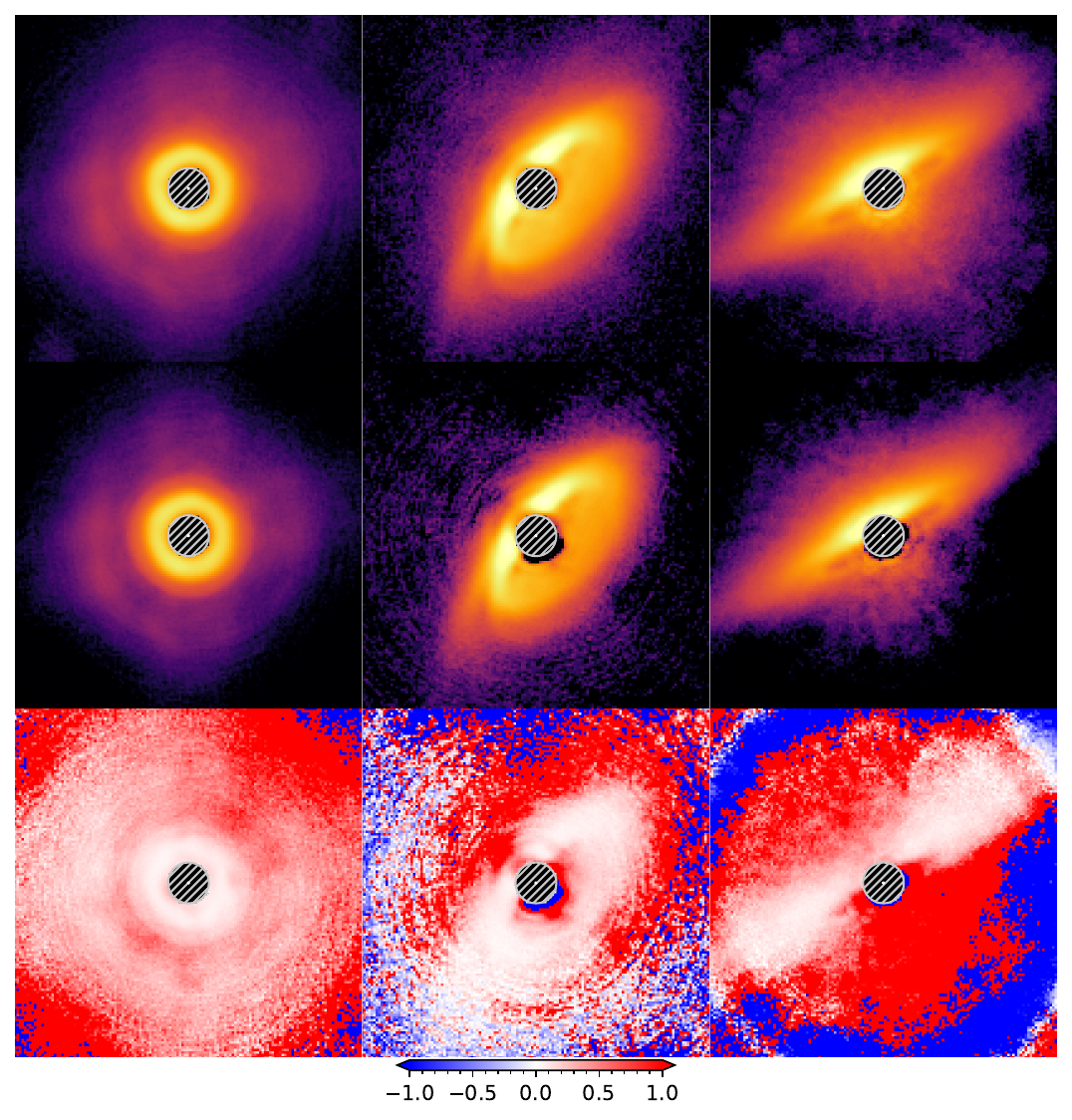}
     \caption{Validation of DIKL results (top) using DI-sNMF results (middle), with the former method being more computationally efficient than the latter by ${\sim}3$ orders of magnitude. The difference between the two methods (bottom), obtained by first subtracting the DI-sNMF result from that of DIKL then divided it by the DI-sNMF result, is ${\lesssim}10\%$ for the disk signals.\\ (The data used to create this figure are available in the ancillary folder on arXiv.)}
     \label{fig-diff}    
 \end{figure}

In comparison with the DI-sNMF method, which is mathematically well-founded to deliver high-quality results \citep[e.g.,][]{ren20, olofsson23, xie23}, DIKL can provide satisfactory results in Fig.~\ref{fig-diff}. The DIKL results not only resemble the morphology of disks from DI-sNMF \citep{olofsson23, ren23}, but also agree with the DI-sNMF surface brightness values within ${\sim}10\%$ for the disk-hosting regions. This further shows that DIKL can be a promising method in reaching high-quality results, and with a computational efficiency that is ${\sim}3$ orders of magnitude better than DI-sNMF.

 With the demonstrated superiority of DIKL over the classical KLIP method in Fig.~\ref{fig-compare}, as well as the consistency of its results with the established DI-sNMF method in Fig.~\ref{fig-diff}, DIKL is a promising method that can produce high-quality results using similar amount of calculation time as KLIP. However, given that the DIKL basis being a non-orthogonal basis in practice, we should not yet ascribe full credibility in the fine details of DIKL results for interpretation until its application to datasets from other instruments is validated. Nevertheless, DIKL possesses a distinctive advantage in producing high-quality preliminary outcomes with high computational efficiency as the first step for data analysis. Such datasets with promising results can be then reduced with more advanced methods including DI-sNMF to ensure signal quality.

\section{Summary}\label{sec-sum}
For RDI data reduction in high-contrast imaging, we demonstrated a high-efficiency and analytical approach to perform data imputation with modified KLIP algorithm: DIKL. Specifically, we can modify two steps in KLIP to reach the purpose of data imputation. On the one hand, in component construction, we use the eigenvectors of the covariance matrix -- from the regions which only host speckle signals (i.e., the anchor matrix, ``a'') -- to generate the DIKL basis for the regions hosting non-speckle signals (i.e., the boat matrix, ``b''), see Eq.~\eqref{eq-kl-transfer}. On the other hand, in speckle removal, we adopt the coefficients from capturing the speckles in the anchor matrix, and apply them to the DIKL basis and thus capture the speckles for the regions that host astrophysical signals in the boat matrix, see Eq.~\eqref{eq-DIKL}. With the two modifications, we only need to perform spectral decomposition once in Eq.~\eqref{eq-decomp}. What is more, the corresponding covariance matrix for the anchor matrix is faster to compute than KLIP, due to a reduced number of matrix elements. As a result, the computational complexity of DIKL is similar as or less than that of KLIP.

By avoiding the projection of astrophysical signals onto speckle features, DIKL can recover face-on structures that are normally overfit in RDI observations, see Fig.~\ref{fig-compare}. On the one hand, in comparison with the mathematically well-founded iterative DI-sNMF method in \citet{ren20}, DIKL might potentially provide analytical results of low quality due to the non-orthogonal DIKL basis in Eq.~\eqref{eq-kl-transfer}. However, DIKL can approach a data quality similar to that of the latter in Fig.~\ref{fig-diff}, since the crosstalk of the DIKL basis is negligible for higher order terms in Figs.~\ref{fig-frv} and \ref{fig-corr-dikl}. On the other hand, in comparison with the \citet{hunziker18} modification of KLIP, DIKL does not apply KL transform for the regions interior to the control ring while the latter does. Given that KL transform captures the variance of the signals \citep[e.g.,][]{soummer12}, while high contrast imaging observations have the highest variance next to the coronagraph \citep[e.g.,][]{pueyo18}, the \citet{hunziker18} and \citet{xuan18} approach should not be adopted to realize the data imputation concept for KLIP. In comparison with that approach, DIKL is not sensitive to such variances, since it uses the control ring signals that are less prone to random noise for KL transform, and thus DIKL is theoretically more plausible for data imputation. As a result, DIKL can be a promising analytical method that provides initial results, and with a ${\sim}3$ orders of magnitude higher computational efficiency than the existing iterative methods \citep[e.g.,][]{ren20}. However, due to the crosstalk of DIKL basis, when the reference images are not stable, we recommend using DIKL for initial signal detection followed by other iterative data imputation methods for detailed characterization.

Given that DIKL is a natural extension of the KLIP algorithm, it can be implemented in the existing high-contrast imaging packages \citep[e.g.,][]{pyklip, vip, pynpoint, adijl} for reference differential imaging data reduction. What is more, DIKL can perform data reduction for images containing negative values (from background removal), and thus it might potentially extract fainter disks than DI-sNMF which only takes in non-negative values. In the upcoming ELT era, we expect DIKL and its future derivative methods to provide high-quality results with high computational efficiency. Moving forward, on the one hand, we can assign different weights to the pixels in KL transform \citep[e.g.,][]{bailey12} to make it less prone to random noise or shot noise. On the other hand, for non-RDI data (e.g., angular differential imaging, spectral differential imaging), more careful derivations which can handle missing data for different regions from different images, including modifying the covariance matrix for KL transform and the KLIP procedure, are needed in the future.

\begin{acknowledgements}
We thank the anonymous referee for their suggestions that increased the readability and clarity of this manuscript, Mamadou N'Diaye for discussions, and Chen Xie for discussions on reference differential imaging in the ELT era. Based on observations collected at the European Organisation for Astronomical Research in the Southern Hemisphere under ESO programmes 105.209E and 105.20GP. We thank the PIs for the two programs (M.~Benisty and J.~Olofsson) for the datasets that validated this study. This research has received funding from the European Union's Horizon 2020 research and innovation programme under the Marie Sk\l odowska-Curie grant agreement No.~101103114. 
\end{acknowledgements}

\bibliography{refs}

\begin{thebibliography}{}
\expandafter\ifx\csname natexlab\endcsname\relax\def\natexlab#1{#1}\fi
\providecommand{\url}[1]{\href{#1}{#1}}
\providecommand{\dodoi}[1]{}
\providecommand{\doarXiv}[1]{\href{https://arxiv.org/abs/#1}{\nolinkurl{https://arxiv.org/abs/#1}}}

\bibitem[{{Amara} \& {Quanz}(2012)}]{amara12}
{Amara}, A., \& {Quanz}, S.~P. 2012,
  \href{http://dx.doi.org/10.1111/j.1365-2966.2012.21918.x}{\color{magenta}\mnras},
  \href{https://ui.adsabs.harvard.edu/abs/2012MNRAS.427..948A}{\color{blue}427},
  \href{https://ui.adsabs.harvard.edu/abs/2012MNRAS.427..948A}{\color{blue}948}

\bibitem[{{Bailey}(2012)}]{bailey12}
{Bailey}, S. 2012,
  \href{http://dx.doi.org/10.1086/668105}{\color{magenta}\pasp},
  \href{https://ui.adsabs.harvard.edu/abs/2012PASP..124.1015B}{\color{blue}124},
  \href{https://ui.adsabs.harvard.edu/abs/2012PASP..124.1015B}{\color{blue}1015}

\bibitem[{{Benisty} {et~al.}(2023){Benisty}, {Dominik}, {Follette}, {Garufi},
  {Ginski}, {Hashimoto}, {Keppler}, {Kley}, \& {Monnier}}]{benisty23}
{Benisty}, M., {Dominik}, C., {Follette}, K., {et~al.} 2023,
  \href{http://dx.doi.org/10.48550/arXiv.2203.09991}{\color{magenta}Astronomical
  Society of the Pacific Conference Series},
  \href{https://ui.adsabs.harvard.edu/abs/2023ASPC..534..605B}{\color{blue}534},
  \href{https://ui.adsabs.harvard.edu/abs/2023ASPC..534..605B}{\color{blue}605}

\bibitem[{{Berdeu} {et~al.}(2022){Berdeu}, {Langlois}, \& {Vachier}}]{berdeu22}
{Berdeu}, A., {Langlois}, M., \& {Vachier}, F. 2022,
  \href{http://dx.doi.org/10.1051/0004-6361/202142623}{\color{magenta}\aap},
  \href{https://ui.adsabs.harvard.edu/abs/2022A&A...658L...4B}{\color{blue}658},
  \href{https://ui.adsabs.harvard.edu/abs/2022A&A...658L...4B}{\color{blue}L4}

\bibitem[{{Bowens} {et~al.}(2021){Bowens}, {Meyer}, {Delacroix}, {Absil}, {van
  Boekel}, {Quanz}, {Shinde}, {Kenworthy}, {Carlomagno}, {Orban de Xivry},
  {Cantalloube}, \& {Pathak}}]{bowens21}
{Bowens}, R., {Meyer}, M.~R., {Delacroix}, C., {et~al.} 2021,
  \href{http://dx.doi.org/10.1051/0004-6361/202141109}{\color{magenta}\aap},
  \href{https://ui.adsabs.harvard.edu/abs/2021A&A...653A...8B}{\color{blue}653},
  \href{https://ui.adsabs.harvard.edu/abs/2021A&A...653A...8B}{\color{blue}A8}

\bibitem[{{Cugno} {et~al.}(2023){Cugno}, {Pearce}, {Launhardt}, {Bonse}, {Ma},
  {Henning}, {Quirrenbach}, {S{\'e}gransan}, {Matthews}, {Quanz}, {Kennedy},
  {M{\"u}ller}, {Reffert}, \& {Rickman}}]{cugno23}
{Cugno}, G., {Pearce}, T.~D., {Launhardt}, R., {et~al.} 2023,
  \href{http://dx.doi.org/10.1051/0004-6361/202244891}{\color{magenta}\aap},
  \href{https://ui.adsabs.harvard.edu/abs/2023A&A...669A.145C}{\color{blue}669},
  \href{https://ui.adsabs.harvard.edu/abs/2023A&A...669A.145C}{\color{blue}A145}

\bibitem[{{Currie} {et~al.}(2023){Currie}, {Biller}, {Lagrange}, {Marois},
  {Guyon}, {Nielsen}, {Bonnefoy}, \& {De Rosa}}]{currie23}
{Currie}, T., {Biller}, B., {Lagrange}, A., {et~al.} 2023,
  \href{http://dx.doi.org/10.48550/arXiv.2205.05696}{\color{magenta}Astronomical
  Society of the Pacific Conference Series},
  \href{https://ui.adsabs.harvard.edu/abs/2023ASPC..534..799C}{\color{blue}534},
  \href{https://ui.adsabs.harvard.edu/abs/2023ASPC..534..799C}{\color{blue}799}

\bibitem[{{Dohlen} {et~al.}(2008){Dohlen}, {Langlois}, {Saisse}, {Hill},
  {Origne}, {Jacquet}, {Fabron}, {Blanc}, {Llored}, {Carle}, {Moutou}, {Vigan},
  {Boccaletti}, {Carbillet}, {Mouillet}, \& {Beuzit}}]{dohlen08}
{Dohlen}, K., {Langlois}, M., {Saisse}, M., {et~al.} 2008,
  \href{http://dx.doi.org/10.1117/12.789786}{\color{magenta}Proc.~SPIE},
  \href{https://ui.adsabs.harvard.edu/abs/2008SPIE.7014E..3LD}{\color{blue}7014},
  \href{https://ui.adsabs.harvard.edu/abs/2008SPIE.7014E..3LD}{\color{blue}70143L}

\bibitem[{{Esposito} {et~al.}(2020){Esposito}, {Kalas}, {Fitzgerald},
  {Millar-Blanchaer}, {Duch{\^e}ne}, {Patience}, {Hom}, {Perrin}, {De Rosa},
  {Chiang}, {Czekala}, {Macintosh}, {Graham}, {Ansdell}, {Arriaga}, {Bruzzone},
  {Bulger}, {Chen}, {Cotten}, {Dong}, {Draper}, {Follette}, {Hung}, {Lopez},
  {Matthews}, {Mazoyer}, {Metchev}, {Rameau}, {Ren}, {Rice}, {Song}, {Stahl},
  {Wang}, {Wolff}, {Zuckerman}, {Ammons}, {Bailey}, {Barman}, {Chilcote},
  {Doyon}, {Gerard}, {Goodsell}, {Greenbaum}, {Hibon}, {Hinkley}, {Ingraham},
  {Konopacky}, {Maire}, {Marchis}, {Marley}, {Marois}, {Nielsen},
  {Oppenheimer}, {Palmer}, {Poyneer}, {Pueyo}, {Rajan}, {Rantakyr{\"o}},
  {Ruffio}, {Savransky}, {Schneider}, {Sivaramakrishnan}, {Soummer}, {Thomas},
  \& {Ward-Duong}}]{esposito20}
{Esposito}, T.~M., {Kalas}, P., {Fitzgerald}, M.~P., {et~al.} 2020,
  \href{http://dx.doi.org/10.3847/1538-3881/ab9199}{\color{magenta}\aj},
  \href{https://ui.adsabs.harvard.edu/abs/2020AJ....160...24E}{\color{blue}160},
  \href{https://ui.adsabs.harvard.edu/abs/2020AJ....160...24E}{\color{blue}24}

\bibitem[{{Flasseur} {et~al.}(2021){Flasseur}, {Th{\'e}}, {Denis},
  {Thi{\'e}baut}, \& {Langlois}}]{flasseur21}
{Flasseur}, O., {Th{\'e}}, S., {Denis}, L., {et~al.} 2021,
  \href{http://dx.doi.org/10.1051/0004-6361/202038957}{\color{magenta}\aap},
  \href{https://ui.adsabs.harvard.edu/abs/2021A&A...651A..62F}{\color{blue}651},
  \href{https://ui.adsabs.harvard.edu/abs/2021A&A...651A..62F}{\color{blue}A62}

\bibitem[{{Follette}(2023)}]{follette23}
{Follette}, K.~B. 2023,
  \href{https://arxiv.org/abs/2308.01354}{\color{magenta}arXiv},
  \href{https://ui.adsabs.harvard.edu/abs/2023arXiv230801354F}{\color{blue}arXiv:2308.01354}

\bibitem[{{Galicher} \& {Marois}(2011)}]{galicher11}
{Galicher}, R., \& {Marois}, C. 2011,
  \href{https://ui.adsabs.harvard.edu/abs/2011aoel.confP..25G}{\color{blue}P25}

\bibitem[{{Gilmozzi} \& {Spyromilio}(2007)}]{elt}
{Gilmozzi}, R., \& {Spyromilio}, J. 2007, The Messenger,
  \href{https://ui.adsabs.harvard.edu/abs/2007Msngr.127...11G}{\color{blue}127},
  \href{https://ui.adsabs.harvard.edu/abs/2007Msngr.127...11G}{\color{blue}11}

\bibitem[{{Ginski} {et~al.}(2021){Ginski}, {Facchini}, {Huang}, {Benisty},
  {Vaendel}, {Stapper}, {Dominik}, {Bae}, {M{\'e}nard}, {Muro-Arena},
  {Hogerheijde}, {McClure}, {van Holstein}, {Birnstiel}, {Boehler}, {Bohn},
  {Flock}, {Mamajek}, {Manara}, {Pinilla}, {Pinte}, \& {Ribas}}]{ginski21}
{Ginski}, C., {Facchini}, S., {Huang}, J., {et~al.} 2021,
  \href{http://dx.doi.org/10.3847/2041-8213/abdf57}{\color{magenta}\apjl},
  \href{https://ui.adsabs.harvard.edu/abs/2021ApJ...908L..25G}{\color{blue}908},
  \href{https://ui.adsabs.harvard.edu/abs/2021ApJ...908L..25G}{\color{blue}L25}

\bibitem[{{Gomez Gonzalez} {et~al.}(2016){Gomez Gonzalez}, {Wertz},
  {Christiaens}, {Absil}, \& {Mawet}}]{vip}
{Gomez Gonzalez}, C.~A., {Wertz}, O., {Christiaens}, V., {et~al.} 2016, {VIP:
  Vortex Image Processing pipeline for high-contrast direct imaging of
  exoplanets}, \href{http://ascl.net/1603.003}{\color{magenta}ASCL},
  \href{https://ui.adsabs.harvard.edu/abs/2016ascl.soft03003G}{\color{blue}1603.003}

\bibitem[{{Gratadour} {et~al.}(2015){Gratadour}, {Rouan}, {Grosset},
  {Boccaletti}, \& {Cl{\'e}net}}]{gratadour15}
{Gratadour}, D., {Rouan}, D., {Grosset}, L., {et~al.} 2015,
  \href{http://dx.doi.org/10.1051/0004-6361/201526554}{\color{magenta}\aap},
  \href{https://ui.adsabs.harvard.edu/abs/2015A&A...581L...8G}{\color{blue}581},
  \href{https://ui.adsabs.harvard.edu/abs/2015A&A...581L...8G}{\color{blue}L8}

\bibitem[{{Harris} {et~al.}(2020){Harris}, {Millman}, {van der Walt},
  {Gommers}, {Virtanen}, {Cournapeau}, {Wieser}, {Taylor}, {Berg}, {Smith},
  {Kern}, {Picus}, {Hoyer}, {van Kerkwijk}, {Brett}, {Haldane}, {del R{\'\i}o},
  {Wiebe}, {Peterson}, {G{\'e}rard-Marchant}, {Sheppard}, {Reddy}, {Weckesser},
  {Abbasi}, {Gohlke}, \& {Oliphant}}]{numpy}
{Harris}, C.~R., {Millman}, K.~J., {van der Walt}, S.~J., {et~al.} 2020,
  \href{http://dx.doi.org/10.1038/s41586-020-2649-2}{\color{magenta}\nat},
  \href{https://ui.adsabs.harvard.edu/abs/2020Natur.585..357H}{\color{blue}585},
  \href{https://ui.adsabs.harvard.edu/abs/2020Natur.585..357H}{\color{blue}357}

\bibitem[{{Hunziker} {et~al.}(2018){Hunziker}, {Quanz}, {Amara}, \&
  {Meyer}}]{hunziker18}
{Hunziker}, S., {Quanz}, S.~P., {Amara}, A., \& {Meyer}, M.~R. 2018,
  \href{http://dx.doi.org/10.1051/0004-6361/201731428}{\color{magenta}\aap},
  \href{https://ui.adsabs.harvard.edu/abs/2018A&A...611A..23H}{\color{blue}611},
  \href{https://ui.adsabs.harvard.edu/abs/2018A&A...611A..23H}{\color{blue}A23}

\bibitem[{{Juillard} {et~al.}(2022){Juillard}, {Christiaens}, \&
  {Absil}}]{juillard22}
{Juillard}, S., {Christiaens}, V., \& {Absil}, O. 2022,
  \href{http://dx.doi.org/10.1051/0004-6361/202244402}{\color{magenta}\aap},
  \href{https://ui.adsabs.harvard.edu/abs/2022A&A...668A.125J}{\color{blue}668},
  \href{https://ui.adsabs.harvard.edu/abs/2022A&A...668A.125J}{\color{blue}A125}

\bibitem[{{Lafreni{\`e}re} {et~al.}(2007){Lafreni{\`e}re}, {Marois}, {Doyon},
  {Nadeau}, \& {Artigau}}]{lafreniere07}
{Lafreni{\`e}re}, D., {Marois}, C., {Doyon}, R., {et~al.} 2007,
  \href{http://dx.doi.org/10.1086/513180}{\color{magenta}\apj},
  \href{https://ui.adsabs.harvard.edu/abs/2007ApJ...660..770L}{\color{blue}660},
  \href{https://ui.adsabs.harvard.edu/abs/2007ApJ...660..770L}{\color{blue}770}

\bibitem[{Lee \& Seung(2001)}]{lee01}
Lee, D.~D., \& Seung, H.~S. 2001, in
  \href{http://papers.nips.cc/paper/1861-algorithms-for-non-negative-matrix-factorization.pdf}{\color{magenta}Advances
  in Neural Information Processing Systems 13}, ed. T.~K. Leen, T.~G.
  Dietterich, \& V.~Tresp (MIT Press), 556

\bibitem[{{Lucas} \& {Bottom}(2020)}]{adijl}
{Lucas}, M., \& {Bottom}, M. 2020,
  \href{http://dx.doi.org/10.21105/joss.02843}{\color{magenta}JOSS},
  \href{https://ui.adsabs.harvard.edu/abs/2020JOSS....5.2843L}{\color{blue}5},
  \href{https://ui.adsabs.harvard.edu/abs/2020JOSS....5.2843L}{\color{blue}2843}

\bibitem[{{Maire} {et~al.}(2016){Maire}, {Langlois}, {Dohlen}, {Lagrange},
  {Gratton}, {Chauvin}, {Desidera}, {Girard}, {Milli}, {Vigan}, {Zins},
  {Delorme}, {Beuzit}, {Claudi}, {Feldt}, {Mouillet}, {Puget}, {Turatto}, \&
  {Wildi}}]{maire16}
{Maire}, A.-L., {Langlois}, M., {Dohlen}, K., {et~al.} 2016,
  \href{http://dx.doi.org/10.1117/12.2233013}{\color{magenta}Proc.~SPIE},
  \href{https://ui.adsabs.harvard.edu/abs/2016SPIE.9908E..34M}{\color{blue}9908},
  \href{https://ui.adsabs.harvard.edu/abs/2016SPIE.9908E..34M}{\color{blue}990834}

\bibitem[{{Marois} {et~al.}(2006){Marois}, {Lafreni{\`e}re}, {Doyon},
  {Macintosh}, \& {Nadeau}}]{marois06}
{Marois}, C., {Lafreni{\`e}re}, D., {Doyon}, R., {et~al.} 2006,
  \href{http://dx.doi.org/10.1086/500401}{\color{magenta}\apj},
  \href{https://ui.adsabs.harvard.edu/abs/2006ApJ...641..556M}{\color{blue}641},
  \href{https://ui.adsabs.harvard.edu/abs/2006ApJ...641..556M}{\color{blue}556}

\bibitem[{{Matthews} {et~al.}(2017){Matthews}, {Hinkley}, {Vigan}, {Kennedy},
  {Rizzuto}, {Stapelfeldt}, {Mawet}, {Booth}, {Chen}, \&
  {Jang-Condell}}]{matthews17}
{Matthews}, E., {Hinkley}, S., {Vigan}, A., {et~al.} 2017,
  \href{http://dx.doi.org/10.3847/2041-8213/aa7943}{\color{magenta}\apjl},
  \href{https://ui.adsabs.harvard.edu/abs/2017ApJ...843L..12M}{\color{blue}843},
  \href{https://ui.adsabs.harvard.edu/abs/2017ApJ...843L..12M}{\color{blue}L12}

\bibitem[{{Mazoyer} {et~al.}(2020){Mazoyer}, {Arriaga}, {Hom},
  {Millar-Blanchaer}, {Chen}, {Wang}, {Duch{\^e}ne}, {Patience}, \&
  {Pueyo}}]{mazoyer20}
{Mazoyer}, J., {Arriaga}, P., {Hom}, J., {et~al.} 2020,
  \href{http://dx.doi.org/10.1117/12.2560091}{\color{magenta}Proc.~SPIE},
  \href{https://ui.adsabs.harvard.edu/abs/2020SPIE11447E..59M}{\color{blue}11447},
  \href{https://ui.adsabs.harvard.edu/abs/2020SPIE11447E..59M}{\color{blue}1144759}

\bibitem[{{Milli} {et~al.}(2012){Milli}, {Mouillet}, {Lagrange}, {Boccaletti},
  {Mawet}, {Chauvin}, \& {Bonnefoy}}]{milli12}
{Milli}, J., {Mouillet}, D., {Lagrange}, A.~M., {et~al.} 2012,
  \href{http://dx.doi.org/10.1051/0004-6361/201219687}{\color{magenta}\aap},
  \href{https://ui.adsabs.harvard.edu/abs/2012A&A...545A.111M}{\color{blue}545},
  \href{https://ui.adsabs.harvard.edu/abs/2012A&A...545A.111M}{\color{blue}A111}

\bibitem[{{Milli} {et~al.}(2017){Milli}, {Vigan}, {Mouillet}, {Lagrange},
  {Augereau}, {Pinte}, {Mawet}, {Schmid}, {Boccaletti}, {Matr{\`a}}, {Kral},
  {Ertel}, {Chauvin}, {Bazzon}, {M{\'e}nard}, {Beuzit}, {Thalmann}, {Dominik},
  {Feldt}, {Henning}, {Min}, {Girard}, {Galicher}, {Bonnefoy}, {Fusco}, {de
  Boer}, {Janson}, {Maire}, {Mesa}, {Schlieder}, \& {SPHERE
  Consortium}}]{milli17}
{Milli}, J., {Vigan}, A., {Mouillet}, D., {et~al.} 2017,
  \href{http://dx.doi.org/10.1051/0004-6361/201527838}{\color{magenta}\aap},
  \href{https://ui.adsabs.harvard.edu/abs/2017A&A...599A.108M}{\color{blue}599},
  \href{https://ui.adsabs.harvard.edu/abs/2017A&A...599A.108M}{\color{blue}A108}

\bibitem[{{Nielsen} {et~al.}(2019){Nielsen}, {De Rosa}, {Macintosh}, {Wang},
  {Ruffio}, {Chiang}, {Marley}, {Saumon}, {Savransky}, {Ammons}, {Bailey},
  {Barman}, {Blain}, {Bulger}, {Burrows}, {Chilcote}, {Cotten}, {Czekala},
  {Doyon}, {Duch{\^e}ne}, {Esposito}, {Fabrycky}, {Fitzgerald}, {Follette},
  {Fortney}, {Gerard}, {Goodsell}, {Graham}, {Greenbaum}, {Hibon}, {Hinkley},
  {Hirsch}, {Hom}, {Hung}, {Dawson}, {Ingraham}, {Kalas}, {Konopacky},
  {Larkin}, {Lee}, {Lin}, {Maire}, {Marchis}, {Marois}, {Metchev},
  {Millar-Blanchaer}, {Morzinski}, {Oppenheimer}, {Palmer}, {Patience},
  {Perrin}, {Poyneer}, {Pueyo}, {Rafikov}, {Rajan}, {Rameau}, {Rantakyr{\"o}},
  {Ren}, {Schneider}, {Sivaramakrishnan}, {Song}, {Soummer}, {Tallis},
  {Thomas}, {Ward-Duong}, \& {Wolff}}]{nielsen19}
{Nielsen}, E.~L., {De Rosa}, R.~J., {Macintosh}, B., {et~al.} 2019,
  \href{http://dx.doi.org/10.3847/1538-3881/ab16e9}{\color{magenta}\aj},
  \href{https://ui.adsabs.harvard.edu/abs/2019AJ....158...13N}{\color{blue}158},
  \href{https://ui.adsabs.harvard.edu/abs/2019AJ....158...13N}{\color{blue}13}

\bibitem[{{Olofsson} {et~al.}(2023){Olofsson}, {Th{\'e}bault}, {Bayo}, {Milli},
  {van Holstein}, {Henning}, {Medina-Olea}, {Godoy}, \&
  {Mauc{\'o}}}]{olofsson23}
{Olofsson}, J., {Th{\'e}bault}, P., {Bayo}, A., {et~al.} 2023,
  \href{http://dx.doi.org/10.1051/0004-6361/202346097}{\color{magenta}\aap},
  \href{https://ui.adsabs.harvard.edu/abs/2023A&A...674A..84O}{\color{blue}674},
  \href{https://ui.adsabs.harvard.edu/abs/2023A&A...674A..84O}{\color{blue}A84}

\bibitem[{{Oppenheimer} \& {Hinkley}(2009)}]{oppenheimer09}
{Oppenheimer}, B.~R., \& {Hinkley}, S. 2009,
  \href{http://dx.doi.org/10.1146/annurev-astro-082708-101717}{\color{magenta}\araa},
  \href{https://ui.adsabs.harvard.edu/abs/2009ARA&A..47..253O}{\color{blue}47},
  \href{https://ui.adsabs.harvard.edu/abs/2009ARA&A..47..253O}{\color{blue}253}

\bibitem[{{Pairet} {et~al.}(2018){Pairet}, {Cantalloube}, \&
  {Jacques}}]{pairet18}
{Pairet}, B., {Cantalloube}, F., \& {Jacques}, L. 2018,
  \href{https://arxiv.org/abs/1812.01333}{\color{magenta}arXiv},
  \href{https://ui.adsabs.harvard.edu/abs/2018arXiv181201333P}{\color{blue}arXiv:1812.01333}

\bibitem[{{Pairet} {et~al.}(2021){Pairet}, {Cantalloube}, \&
  {Jacques}}]{pairet21}
---. 2021,
  \href{http://dx.doi.org/10.1093/mnras/stab607}{\color{magenta}\mnras},
  \href{https://ui.adsabs.harvard.edu/abs/2021MNRAS.503.3724P}{\color{blue}503},
  \href{https://ui.adsabs.harvard.edu/abs/2021MNRAS.503.3724P}{\color{blue}3724}

\bibitem[{{Perrin} {et~al.}(2015){Perrin}, {Duchene}, {Millar-Blanchaer},
  {Fitzgerald}, {Graham}, {Wiktorowicz}, {Kalas}, {Macintosh}, {Bauman},
  {Cardwell}, {Chilcote}, {De Rosa}, {Dillon}, {Doyon}, {Dunn}, {Erikson},
  {Gavel}, {Goodsell}, {Hartung}, {Hibon}, {Ingraham}, {Kerley}, {Konapacky},
  {Larkin}, {Maire}, {Marchis}, {Marois}, {Mittal}, {Morzinski}, {Oppenheimer},
  {Palmer}, {Patience}, {Poyneer}, {Pueyo}, {Rantakyr{\"o}}, {Sadakuni},
  {Saddlemyer}, {Savransky}, {Soummer}, {Sivaramakrishnan}, {Song}, {Thomas},
  {Wallace}, {Wang}, \& {Wolff}}]{perrin15}
{Perrin}, M.~D., {Duchene}, G., {Millar-Blanchaer}, M., {et~al.} 2015,
  \href{http://dx.doi.org/10.1088/0004-637X/799/2/182}{\color{magenta}\apj},
  \href{https://ui.adsabs.harvard.edu/abs/2015ApJ...799..182P}{\color{blue}799},
  \href{https://ui.adsabs.harvard.edu/abs/2015ApJ...799..182P}{\color{blue}182}

\bibitem[{{Pohl} {et~al.}(2017){Pohl}, {Benisty}, {Pinilla}, {Ginski}, {de
  Boer}, {Avenhaus}, {Henning}, {Zurlo}, {Boccaletti}, {Augereau}, {Birnstiel},
  {Dominik}, {Facchini}, {Fedele}, {Janson}, {Keppler}, {Kral}, {Langlois},
  {Ligi}, {Maire}, {M{\'e}nard}, {Meyer}, {Pinte}, {Quanz}, {Sauvage},
  {Sezestre}, {Stolker}, {Szul{\'a}gyi}, {van Boekel}, {van der Plas},
  {Villenave}, {Baruffolo}, {Baudoz}, {Le Mignant}, {Maurel}, {Ramos}, \&
  {Weber}}]{pohl17}
{Pohl}, A., {Benisty}, M., {Pinilla}, P., {et~al.} 2017,
  \href{http://dx.doi.org/10.3847/1538-4357/aa94c2}{\color{magenta}\apj},
  \href{https://ui.adsabs.harvard.edu/abs/2017ApJ...850...52P}{\color{blue}850},
  \href{https://ui.adsabs.harvard.edu/abs/2017ApJ...850...52P}{\color{blue}52}

\bibitem[{{Pueyo}(2016)}]{pueyo16}
{Pueyo}, L. 2016,
  \href{http://dx.doi.org/10.3847/0004-637X/824/2/117}{\color{magenta}\apj},
  \href{https://ui.adsabs.harvard.edu/abs/2016ApJ...824..117P}{\color{blue}824},
  \href{https://ui.adsabs.harvard.edu/abs/2016ApJ...824..117P}{\color{blue}117}

\bibitem[{{Pueyo}(2018)}]{pueyo18}
---. 2018, {Direct Imaging as a Detection Technique for Exoplanets},
  \href{http://adsabs.harvard.edu/abs/2018haex.bookE..10P}{\color{blue}10}

\bibitem[{{Ren} {et~al.}(2023){Ren}, {Benisty}, {Ginksi}, {et~al.}}]{ren23}
{Ren}, B., {Benisty}, M., {Ginksi}, C., {et~al.} 2023, \aap, under review

\bibitem[{{Ren} {et~al.}(2020){Ren}, {Pueyo}, {Chen}, {Choquet}, {Debes},
  {Duch{\^e}ne}, {M{\'e}nard}, \& {Perrin}}]{ren20}
{Ren}, B., {Pueyo}, L., {Chen}, C., {et~al.} 2020,
  \href{http://dx.doi.org/10.3847/1538-4357/ab7024}{\color{magenta}\apj},
  \href{https://ui.adsabs.harvard.edu/abs/2020arXiv200100563R}{\color{blue}892},
  \href{https://ui.adsabs.harvard.edu/abs/2020arXiv200100563R}{\color{blue}74}

\bibitem[{{Ren} {et~al.}(2018){Ren}, {Pueyo}, {Zhu}, {Debes}, \&
  {Duch{\^e}ne}}]{ren18}
{Ren}, B., {Pueyo}, L., {Zhu}, G.~B., {et~al.} 2018,
  \href{http://dx.doi.org/10.3847/1538-4357/aaa1f2}{\color{magenta}\apj},
  \href{https://ui.adsabs.harvard.edu/abs/2018ApJ...852..104R}{\color{blue}852},
  \href{https://ui.adsabs.harvard.edu/abs/2018ApJ...852..104R}{\color{blue}104}

\bibitem[{{Ruane} {et~al.}(2019){Ruane}, {Ngo}, {Mawet}, {Absil}, {Choquet},
  {Cook}, {Gomez Gonzalez}, {Huby}, {Matthews}, {Meshkat}, {Reggiani},
  {Serabyn}, {Wallack}, \& {Xuan}}]{ruane19}
{Ruane}, G., {Ngo}, H., {Mawet}, D., {et~al.} 2019,
  \href{http://dx.doi.org/10.3847/1538-3881/aafee2}{\color{magenta}\aj},
  \href{https://ui.adsabs.harvard.edu/abs/2019AJ....157..118R}{\color{blue}157},
  \href{https://ui.adsabs.harvard.edu/abs/2019AJ....157..118R}{\color{blue}118}

\bibitem[{{Samland} {et~al.}(2021){Samland}, {Bouwman}, {Hogg}, {Brandner},
  {Henning}, \& {Janson}}]{samland21}
{Samland}, M., {Bouwman}, J., {Hogg}, D.~W., {et~al.} 2021,
  \href{http://dx.doi.org/10.1051/0004-6361/201937308}{\color{magenta}\aap},
  \href{https://ui.adsabs.harvard.edu/abs/2021A&A...646A..24S}{\color{blue}646},
  \href{https://ui.adsabs.harvard.edu/abs/2021A&A...646A..24S}{\color{blue}A24}

\bibitem[{{Soummer} {et~al.}(2012){Soummer}, {Pueyo}, \& {Larkin}}]{soummer12}
{Soummer}, R., {Pueyo}, L., \& {Larkin}, J. 2012,
  \href{http://dx.doi.org/10.1088/2041-8205/755/2/L28}{\color{magenta}\apjl},
  \href{https://ui.adsabs.harvard.edu/abs/2012ApJ...755L..28S}{\color{blue}755},
  \href{https://ui.adsabs.harvard.edu/abs/2012ApJ...755L..28S}{\color{blue}L28}

\bibitem[{{Stapper} \& {Ginski}(2022)}]{stapper22}
{Stapper}, L.~M., \& {Ginski}, C. 2022,
  \href{http://dx.doi.org/10.1051/0004-6361/202142820}{\color{magenta}\aap},
  \href{https://ui.adsabs.harvard.edu/abs/2022A&A...668A..50S}{\color{blue}668},
  \href{https://ui.adsabs.harvard.edu/abs/2022A&A...668A..50S}{\color{blue}A50}

\bibitem[{{Stolker} {et~al.}(2019){Stolker}, {Bonse}, {Quanz}, {Amara},
  {Cugno}, {Bohn}, \& {Boehle}}]{pynpoint}
{Stolker}, T., {Bonse}, M.~J., {Quanz}, S.~P., {et~al.} 2019,
  \href{http://dx.doi.org/10.1051/0004-6361/201834136}{\color{magenta}\aap},
  \href{https://ui.adsabs.harvard.edu/abs/2019A&A...621A..59S}{\color{blue}621},
  \href{https://ui.adsabs.harvard.edu/abs/2019A&A...621A..59S}{\color{blue}A59}

\bibitem[{{van Holstein} {et~al.}(2017){van Holstein}, {Snik}, {Girard}, {de
  Boer}, {Ginski}, {Keller}, {Stam}, {Beuzit}, {Mouillet}, {Kasper},
  {Langlois}, {Zurlo}, {de Kok}, \& {Vigan}}]{irdap1}
{van Holstein}, R.~G., {Snik}, F., {Girard}, J.~H., {et~al.} 2017,
  \href{http://dx.doi.org/10.1117/12.2272554}{\color{magenta}Proc.~SPIE},
  \href{https://ui.adsabs.harvard.edu/abs/2017SPIE10400E..15V}{\color{blue}10400},
  \href{https://ui.adsabs.harvard.edu/abs/2017SPIE10400E..15V}{\color{blue}1040015}

\bibitem[{{van Holstein} {et~al.}(2020){van Holstein}, {Girard}, {de Boer},
  {Snik}, {Milli}, {Stam}, {Ginski}, {Mouillet}, {Wahhaj}, {Schmid}, {Keller},
  {Langlois}, {Dohlen}, {Vigan}, {Pohl}, {Carbillet}, {Fantinel}, {Maurel},
  {Orign{\'e}}, {Petit}, {Ramos}, {Rigal}, {Sevin}, {Boccaletti}, {Le
  Coroller}, {Dominik}, {Henning}, {Lagadec}, {M{\'e}nard}, {Turatto}, {Udry},
  {Chauvin}, {Feldt}, \& {Beuzit}}]{irdap2}
{van Holstein}, R.~G., {Girard}, J.~H., {de Boer}, J., {et~al.} 2020,
  \href{http://dx.doi.org/10.1051/0004-6361/201834996}{\color{magenta}\aap},
  \href{https://ui.adsabs.harvard.edu/abs/2020A&A...633A..64V}{\color{blue}633},
  \href{https://ui.adsabs.harvard.edu/abs/2020A&A...633A..64V}{\color{blue}A64}

\bibitem[{{Vigan} {et~al.}(2021){Vigan}, {Fontanive}, {Meyer}, {Biller},
  {Bonavita}, {Feldt}, {Desidera}, {Marleau}, {Emsenhuber}, {Galicher}, {Rice},
  {Forgan}, {Mordasini}, {Gratton}, {Le Coroller}, {Maire}, {Cantalloube},
  {Chauvin}, {Cheetham}, {Hagelberg}, {Lagrange}, {Langlois}, {Bonnefoy},
  {Beuzit}, {Boccaletti}, {D'Orazi}, {Delorme}, {Dominik}, {Henning}, {Janson},
  {Lagadec}, {Lazzoni}, {Ligi}, {Menard}, {Mesa}, {Messina}, {Moutou},
  {M{\"u}ller}, {Perrot}, {Samland}, {Schmid}, {Schmidt}, {Sissa}, {Turatto},
  {Udry}, {Zurlo}, {Abe}, {Antichi}, {Asensio-Torres}, {Baruffolo}, {Baudoz},
  {Baudrand}, {Bazzon}, {Blanchard}, {Bohn}, {Brown Sevilla}, {Carbillet},
  {Carle}, {Cascone}, {Charton}, {Claudi}, {Costille}, {De Caprio},
  {Delboulb{\'e}}, {Dohlen}, {Engler}, {Fantinel}, {Feautrier}, {Fusco},
  {Gigan}, {Girard}, {Giro}, {Gisler}, {Gluck}, {Gry}, {Hubin}, {Hugot},
  {Jaquet}, {Kasper}, {Le Mignant}, {Llored}, {Madec}, {Magnard}, {Martinez},
  {Maurel}, {M{\"o}ller-Nilsson}, {Mouillet}, {Moulin}, {Orign{\'e}}, {Pavlov},
  {Perret}, {Petit}, {Pragt}, {Puget}, {Rabou}, {Ramos}, {Rickman}, {Rigal},
  {Rochat}, {Roelfsema}, {Rousset}, {Roux}, {Salasnich}, {Sauvage}, {Sevin},
  {Soenke}, {Stadler}, {Suarez}, {Wahhaj}, {Weber}, \& {Wildi}}]{vigan21}
{Vigan}, A., {Fontanive}, C., {Meyer}, M., {et~al.} 2021,
  \href{http://dx.doi.org/10.1051/0004-6361/202038107}{\color{magenta}\aap},
  \href{https://ui.adsabs.harvard.edu/abs/2021A&A...651A..72V}{\color{blue}651},
  \href{https://ui.adsabs.harvard.edu/abs/2021A&A...651A..72V}{\color{blue}A72}

\bibitem[{{Wagner} {et~al.}(2020){Wagner}, {Stone}, {Dong}, {Ertel}, {Apai},
  {Doelman}, {Bohn}, {Najita}, {Brittain}, {Kenworthy}, {Keppler}, {Webster},
  {Mailhot}, \& {Snik}}]{wagner20}
{Wagner}, K., {Stone}, J., {Dong}, R., {et~al.} 2020,
  \href{http://dx.doi.org/10.3847/1538-3881/ab893f}{\color{magenta}\aj},
  \href{https://ui.adsabs.harvard.edu/abs/2020AJ....159..252W}{\color{blue}159},
  \href{https://ui.adsabs.harvard.edu/abs/2020AJ....159..252W}{\color{blue}252}

\bibitem[{{Wahhaj} {et~al.}(2021){Wahhaj}, {Milli}, {Romero}, {Cieza}, {Zurlo},
  {Vigan}, {Pe{\~n}a}, {Valdes}, {Cantalloube}, {Girard}, \&
  {Pantoja}}]{wahhaj21}
{Wahhaj}, Z., {Milli}, J., {Romero}, C., {et~al.} 2021,
  \href{http://dx.doi.org/10.1051/0004-6361/202038794}{\color{magenta}\aap},
  \href{https://ui.adsabs.harvard.edu/abs/2021A&A...648A..26W}{\color{blue}648},
  \href{https://ui.adsabs.harvard.edu/abs/2021A&A...648A..26W}{\color{blue}A26}

\bibitem[{{Wallack} {et~al.}(2023){Wallack}, {Ruffio}, {Ruane},
  {et~al.}}]{wallack23}
{Wallack}, N.~L., {Ruffio}, J.-B., {Ruane}, G., {et~al.} 2023, \aj, accepted

\bibitem[{{Wang} {et~al.}(2015){Wang}, {Ruffio}, {De Rosa}, {Aguilar}, {Wolff},
  \& {Pueyo}}]{pyklip}
{Wang}, J.~J., {Ruffio}, J.-B., {De Rosa}, R.~J., {et~al.} 2015, {pyKLIP: PSF
  Subtraction for Exoplanets and Disks},
  \href{http://ascl.net/1506.001}{\color{magenta}ASCL},
  \href{https://ui.adsabs.harvard.edu/abs/2015ascl.soft06001W}{\color{blue}1506.001}

\bibitem[{{Xie} {et~al.}(2022){Xie}, {Choquet}, {Vigan}, {Cantalloube},
  {Benisty}, {Boccaletti}, {Bonnefoy}, {Desgrange}, {Garufi}, {Girard},
  {Hagelberg}, {Janson}, {Kenworthy}, {Lagrange}, {Langlois}, {Menard}, \&
  {Zurlo}}]{xie22}
{Xie}, C., {Choquet}, E., {Vigan}, A., {et~al.} 2022,
  \href{http://dx.doi.org/10.1051/0004-6361/202243379}{\color{magenta}\aap},
  \href{https://ui.adsabs.harvard.edu/abs/2022A&A...666A..32X}{\color{blue}666},
  \href{https://ui.adsabs.harvard.edu/abs/2022A&A...666A..32X}{\color{blue}A32}

\bibitem[{{Xie} {et~al.}(2023){Xie}, {Ren}, {Dong}, {Choquet}, {Vigan},
  {Gonzalez}, {Wagner}, {Fang}, \& {Ubeira-Gabellini}}]{xie23}
{Xie}, C., {Ren}, B.~B., {Dong}, R., {et~al.} 2023,
  \href{http://dx.doi.org/10.1051/0004-6361/202346305}{\color{magenta}\aap},
  \href{https://ui.adsabs.harvard.edu/abs/2023A&A...675L...1X}{\color{blue}675},
  \href{https://ui.adsabs.harvard.edu/abs/2023A&A...675L...1X}{\color{blue}L1}

\bibitem[{{Xuan} {et~al.}(2018){Xuan}, {Mawet}, {Ngo}, {Ruane}, {Bailey},
  {Choquet}, {Absil}, {Alvarez}, {Bryan}, {Cook}, {Femen{\'\i}a Castell{\'a}},
  {Gomez Gonzalez}, {Huby}, {Knutson}, {Matthews}, {Ragland}, {Serabyn}, \&
  {Zawol}}]{xuan18}
{Xuan}, W.~J., {Mawet}, D., {Ngo}, H., {et~al.} 2018,
  \href{http://dx.doi.org/10.3847/1538-3881/aadae6}{\color{magenta}\aj},
  \href{https://ui.adsabs.harvard.edu/abs/2018AJ....156..156X}{\color{blue}156},
  \href{https://ui.adsabs.harvard.edu/abs/2018AJ....156..156X}{\color{blue}156}

\end{thebibliography}

\appendix
\section{Pseudocode for DIKL implementation}\label{sec-app-pseudo}
We present a pseudocode to implement DIKL in Algorithm~\ref{algo-dikl}. Specifically, we use the KL projection coefficient from Eq.~\eqref{eq-kl-coef}, and apply it to the DIKL basis in Eq.~\eqref{eq-kl-transfer}, to obtain the DIKL projection. We then remove the DIKL projection from the target image to reveal the astrophysical signals in Eq.~\eqref{eq-DIKL}. 

To implement a standalone DIKL function, we need two matrix operations -- matrix multiplication and eigendecomposition -- both are available in modern scientific programming languages. Alternatively, to implement DIKL in the field of high contrast imaging, we can use existing KLIP frameworks (e.g., \texttt{pyKLIP}
: \citealp{pyklip}, \texttt{VIP}
: \citealp{vip}, \texttt{Pynpoint}
: \citealp{pynpoint}, \texttt{ADI.jl}
: \citealp{adijl}), since there are only two additional commands (at the end of Algorithm~\ref{algo-dikl}) in addition to KLIP for RDI observations. There is one extra command to select specific regions to obtain the anchor and boat matrices, and this selection is available using the mask commands in existing frameworks. 

\begin{algorithm}[htb!]
\caption{DIKL algorithm.}
  \begin{algorithmic}[1]
\State  \textbf{Input}: reference array $R$, target vector $t$, anchor selection mask $S^{(a)}$, and boat selection mask $S^{(b)}$; 
\State  Generate anchors: compute anchor reference $R^{(a)}$ and anchor target $t^{(a)}$ using selection mask $S^{(a)}$;  \Comment{See Eq.~\eqref{eq-rab}.}
\State  Generate boats: compute boat reference $R^{(b)}$ and boat target $t^{(b)}$ using selection mask $S^{(b)}$;  \Comment{See Eq.~\eqref{eq-rab}.}
\State $Q^{(a)} \gets$ eigenvectors of $R^{(a)\top}R^{(a)}$; \Comment{See Eq.~\eqref{eq-decomp}.}
\State $Z^{(a)} \gets R^{(a)\top}Q^{(a)}$; \Comment{KL transform in Eq.~\eqref{eq-kl}.}
\State ${c}^{(a)} \gets t^{(a)\top}Z^{(a)}$; \Comment{KL projection in Eq.~\eqref{eq-kl-coef}.}
\State ${Z'}^{(b)} \gets R^{(b)\top}Q^{(a)}$; \Comment{DIKL transform in Eq.~\eqref{eq-kl-transfer}.}
\State $r^{(b)} \gets t^{(b)} - c^{(a)} {Z'}^{(b)}$; \Comment{DIKL projection in Eq.~\eqref{eq-DIKL}.}
\State \textbf{Output}: DIKL residual $r^{(b)}$.
\end{algorithmic}
\label{algo-dikl}
\end{algorithm}

In actual implementation, we recommend splitting Algorithm~\ref{algo-dikl} into two functions for computational efficiency. One function performs KL and DIKL transforms, and returns both the KL basis from Eq.~\eqref{eq-kl} and the DIKL basis from Eq.~\eqref{eq-kl-transfer}. The other uses the KL basis to generate the KL projection coefficients from Eq.~\eqref{eq-kl-coef} for target $t$, then apply the coefficients to the DIKL basis to obtain the residuals in Eq.~\eqref{eq-DIKL}. In this way, we can use the same reference array $R$ for the RDI reduction of different targets using DIKL.

To implement DIKL on existing high contrast imaging pipelines, we detail the key modifications here. Specifically, in the KL transform we need the eigenvector matrix $Q^{(a)}$ in Eq.~\eqref{eq-decomp}, and thus such an output is needed when eigendecomposition is conducted (e.g., \texttt{pyKLIP}\footnote{\url{https://bitbucket.org/pyKLIP/pyklip/src/ab0040da1ae442dce9503502fc92f1736615b37d/pyklip/klip.py\#lines-137}}, \texttt{Pynpoint}\footnote{\url{https://github.com/PynPoint/PynPoint/blob/main/pynpoint/util/psf.py\#L86}}, and \texttt{VIP}\footnote{\url{https://github.com/vortex-exoplanet/VIP/blob/v1.4.0/vip_hci/psfsub/svd.py\#L442C21}}); for \texttt{adi.jl}\footnote{\url{https://github.com/JuliaHCI/ADI.jl/blob/main/src/pca.jl\#L46}} where singular value decomposition is conducted, we can use the output directly. We can then multiply the boat reference matrix $R^{(b)}$ with the eigenvector matrix to obtain the DIKL basis ${Z'}^{(b)}$ in Eq.~\eqref{eq-kl-transfer}. At last, we multiply the KL projection coefficients $c^{(a)}$ from the pipelines for Eq.~\eqref{eq-kl-coef} with the DIKL basis ${Z'}^{(b)}$, then obtain the residuals in Eq.~\eqref{eq-DIKL} after DIKL projection. 

\end{CJK*}
\end{document}